# Relaxation dynamics and colossal magnetocapacitive effect in CdCr$_2$S$_4$

P. Lunkenheimer,[1] R. Fichtl,[1] J. Hemberger,[1] V. Tsurkan,[1,2] and A. Loidl[1]

[1]*Experimental Physics V, Center for Electronic Correlations and Magnetism, University of Augsburg, D-86135 Augsburg, Germany*
[2]*Institute for Applied Physics, Academy of Sciences of Moldova, MD-2028 Chisinau, R. Moldova*

A thorough investigation of the relaxational dynamics in the recently discovered multiferroic CdCr$_2$S$_4$ showing a colossal magnetocapacitive effect has been performed. Broadband dielectric measurements without and with external magnetic fields up to 10 T provide clear evidence that the observed magnetocapacitive effect stems from enormous changes of the relaxation dynamics induced by the development of magnetic order.

Multiferroic materials, showing the simultaneous occurrence of dielectric and magnetic order, are a "hot" topic in recent solid state research [1,2]. Especially the strong variation of electric (magnetic) properties under application of a magnetic (electric) field found in some of these materials, make them highly attractive not only from an academic point of view, but also for potential applications in microelectronics. Among the recently discovered multiferroics, the spinel compound CdCr$_2$S$_4$ is exceptional, showing ferromagnetic and relaxor ferroelectric behavior with sizable ordering temperatures and, most remarkably, a colossal magnetocapacitive (MC) effect of nearly 500% in an external magnetic field $H = 5$ T [2]. However, so far the origin of the strong MC coupling in this compound is unknown. In [2], it was suggested that it may arise from the magnetic field affecting the relaxor dynamics of the polar moments, but alternatively also effects resulting from the hopping dynamics of localized charge carriers were discussed. In the present letter, from broadband dielectric measurements, we provide detailed information on the relaxation dynamics in CdCr$_2$S$_4$, both in the paramagnetic and ferromagnetic state. The results clearly reveal that it is the radical change of the relaxational dynamics driven by the magnetization that leads to the observed colossal MC effects in this compound.

For the present investigation, CdCr$_2$S$_4$ single crystals of the same batch as in [2] were used, with sputtered gold-contacts being applied on opposite sides of the plate-like samples. Dielectric constant and loss were measured over a broad frequency range of 12 decades (3 mHz < $\nu$ < 3 GHz) using frequency-response analysis and a reflectometric technique [3]. A conventional $^4$He bath-cryostat and a cryomagnet allowed measurements at temperatures down to 1.5 K and in magnetic fields up to 10 T. Due to the long data acquisition times at low frequencies, the measurements at $\nu$ < 1 Hz were performed in a restricted temperature range of 50 K - 200 K only, using a closed-cycle refrigerator system.

Figure 1(a) shows the temperature-dependent dielectric constant $\varepsilon'$ of CdCr$_2$S$_4$ for frequencies from 3 mHz to 3 GHz. For $T$ > 100 K, a peak shifting to lower temperatures and increasing in amplitude with decreasing frequency shows up. The dashed line indicates a Curie-Weiss law, 3600 K / ($T$ - 130 K), for the right flank of the peaks, which can be taken as an estimate of the static dielectric constant [4]. Thus the overall characteristics of the relaxational behavior in CdCr$_2$S$_4$ resembles that observed in the so-called relaxor ferroelectrics [5]. There the reduction of $\varepsilon'$ below the peak temperature is usually ascribed to a freezing-in of ferroelectric clusters on the time scale given by the frequency of the applied AC electric-field, quite in contrast to canonical ferroelectrics where the frequency dependence of $\varepsilon'$ is negligible. Concerning the origin of the polar moments in CdCr$_2$S$_4$, in [2] arguments were put forward that ferroelectricity in CdCr$_2$S$_4$ results from an off-center position of the Cr$^{3+}$-ions and that geometrical frustration drives the observed relaxor-like freezing.

The most remarkable feature in Fig. 1(a) is the strong increase of $\varepsilon'(T)$ below the ferromagnetic transition temperature, $T_c = 85$ K, indicating the close coupling of magnetic and dielectric properties [2]. To shed light on the role of the relaxation dynamics in this coupling, in Fig. 1(b) the conductivity $\sigma'(T)$ is shown for various frequencies. As $\sigma' \propto \varepsilon'' \times \nu$, the temperature-dependence of $\sigma'(T)$ is identical to that of the dielectric loss $\varepsilon''$ (we chose a plot of $\sigma'(T)$, to avoid a crossing of the curves obscuring the readability of the figure). Relaxational features seen in $\varepsilon'(T)$ should be accompanied by peaks in $\sigma'(T)$. At $T > T_c$ in CdCr$_2$S$_4$, they should appear at the frequency of the point of inflection at the left wing of the maxima in $\varepsilon'(T)$. However, no such peaks become obvious in Fig. 1(b), which can be ascribed to superimposing contributions from charge carrier transport. In Fig. 1(b), the curve at 3 mHz gives a good approximation of the dc-conductivity, the higher-frequency curves successively branching off from this curve. Subtracting this dc contribution from the higher-frequency curves indeed reveals the expected peaks in $\sigma'(T)$ as demonstrated for 9.5 Hz by the closed squares. Below $T_c$, $\sigma'(T)$ exhibits a rather complex behavior. While there is the general tendency of $\sigma'$ to increase when the sample becomes ferromagnetic and to decrease again at the lowest temperatures, in addition a number of peaks and shoulders appear. The two main peaks [arrows in Fig. 1(b)], shifting with frequency in opposite directions, already indicate that the behavior below $T_c$ is dominated by a complex relaxational behavior as will be discussed in detail below.

In Fig. 2, the influence of an external magnetic field up to 10 T on the dielectric response of CdCr$_2$S$_4$ is demonstrated for two frequencies. A huge increase of $\varepsilon'$ and $\sigma'$ with magnetic field shows up. The amplitude of the MC effect for 9.5 Hz and 5 T, measured by the relative variation of $\varepsilon'$ with field, at $T_c$ amounts to nearly 500% [2] and for 10 T even reaches 3000% [inset of Fig. 2(a)], which to our knowledge is the highest value reported in any material so far. The peaks in $\sigma'(T)$ [Fig. 2(b)] are also strongly affected by the magnetic field, indicating its strong influence on the relaxational dynamics in $CdCr_2S_4$. In addition to the relaxational contribution, $\sigma'(T,H)$ certainly partly is also influenced by the well-known strong magnetoresistance of $CdCr_2S_4$ leading to an increase of the dc conductivity with $H$.

accompanied by peaks in $\varepsilon''(\nu)$. As the peak frequency is related by $\nu_p \approx 1/(2\pi\tau)$ to the relaxation time $\tau$, their shifting to low frequencies with decreasing temperatures reflects the slowing down of the relaxational dynamics as expected for thermally activated processes. The deviations at low frequencies from a simple step or peak function can be ascribed to charge-transport contributions. The shoulder observed in the loss at high frequencies (e.g., at about $10^4$ Hz for 4.2 K) can only be explained by a weak additional relaxation process of so far unknown origin. It leads to additional peaks in $\varepsilon''(T)$ [or $\sigma'(T)$, see Fig. 1(b)], too.

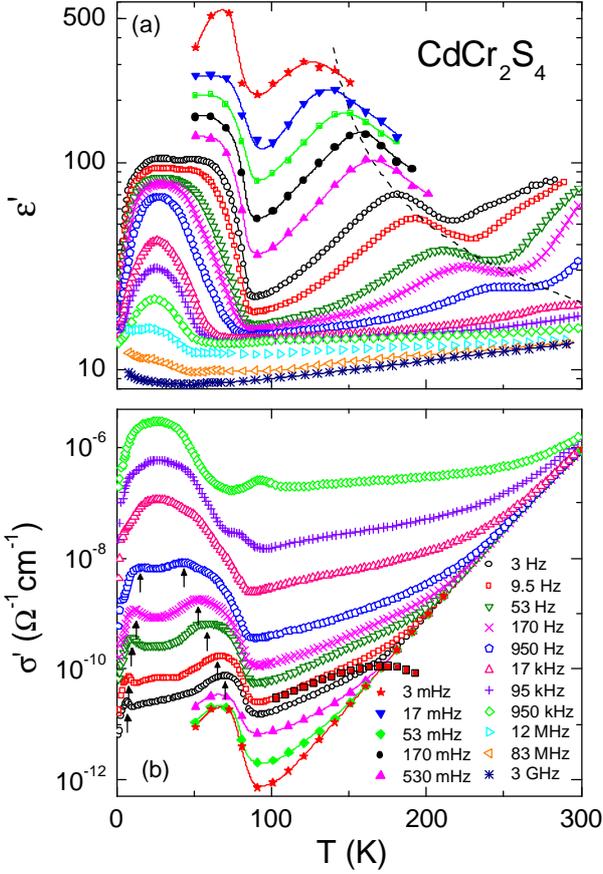

FIG. 1. Temperature dependence of $\varepsilon'$ (a) and $\sigma'$ (b) of $CdCr_2S_4$ for various frequencies [in (b) not all frequencies are shown]. The solid lines are drawn to guide the eyes. The dashed line in (a) indicates the static dielectric susceptibility following a Curie-Weiss like law with a characteristic temperature of 130 K. The closed squares in (b) show $\sigma'(\nu)$ at 9.5 Hz after subtraction of the dc conductivity. The arrows in (b) indicate the relaxation peaks below $T_c$.

The most significant information on relaxational dynamics can be gained from frequency-dependent plots of the permittivity. As was shown in [2], in the temperature region above $T_c$, $\varepsilon''(\nu)$ exhibits the typical peaks shifting towards low temperatures with decreasing frequency that are the signature of relaxational freezing [3,5]. In Figs. 3(a)-(d), $\varepsilon'(\nu)$ and $\varepsilon''(\nu)$ at temperatures below $T_c$ are presented. At $T \leq 28$ K [Figs. 3(a) and (b)], conventional relaxational behavior is observed, namely a steplike decrease of $\varepsilon'(\nu)$,

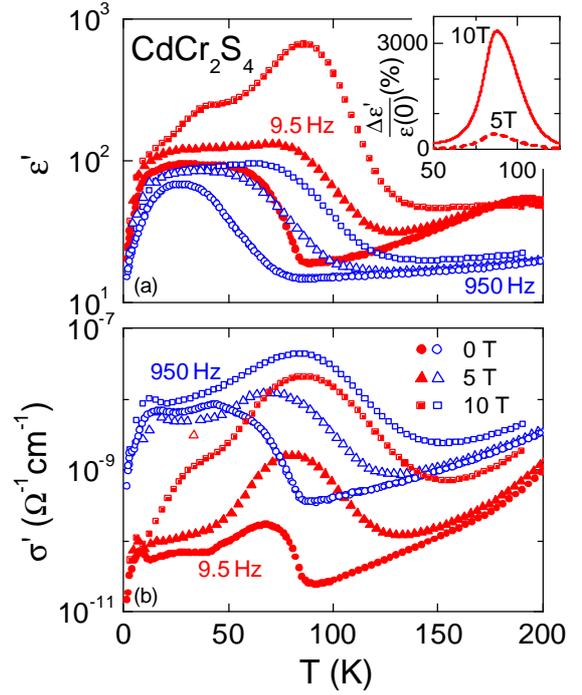

FIG. 2. Temperature dependence of $\varepsilon'$ (a) and $\sigma'$ (b) of $CdCr_2S_4$ with external magnetic fields of 0, 5, and 10 T for two frequencies. The inset provides a measure of the MC effect, with $\Delta\varepsilon' = \varepsilon'(H) - \varepsilon'(0T)$, for 9.5 Hz and magnetic fields H = 5 T and 10 T.

As revealed by Figs. 3(c) and (d), at temperatures between 28 and 81 K, the situation is different: still typical relaxation features show up, however now shifting towards higher frequencies with decreasing temperature! This unexpected behavior signals that the dipolar relaxation rate *in*creases on *de*creasing temperature. It can only be understood assuming that the onset of magnetic order below $T_c$ leads to an acceleration of the relaxation dynamics. This result proves that the strong increase in $\varepsilon'(T)$ below $T_c$ is caused by the speeding up of the relaxation dynamics. Probably the rising of the spin order softens the lattice thereby enhancing the relaxation rate. Also information on the origin of the colossal MC effect in $CdCr_2S_4$ can be deduced from the frequency-dependent results: In Figs. 3(e) and (f), $\varepsilon'(\nu)$ and $\varepsilon''(\nu)$ are shown for two temperatures without and with an applied external magnetic field of 10 T. From the shift of the relaxation features towards higher frequencies, it becomes obvious that an external magnetic field also enhances the relaxation rate.

For a quantitative evaluation of the results of Fig. 3, we fitted the spectra using the Cole-Cole (CC) function, often employed to describe relaxation loss peaks, and a conductivity contribution, $\sigma' = \sigma_{dc} + \sigma_0 \nu^s$, to account for the observed low-frequency upturn of $\varepsilon'' \propto \sigma'/\nu$ [6]. The latter is composed of a dc component and an ac power-law contribution with exponent $s < 1$, representing the so-called "Universal Dielectric Response" (UDR), which in semiconducting systems is the signature of hopping conduction of localized charge carriers [6]. Finally, for the lowest temperatures an additional CC function was assumed, to fit the weak second relaxation process. Good agreement of fits (lines in Fig. 3) and experimental spectra is achieved in this way.

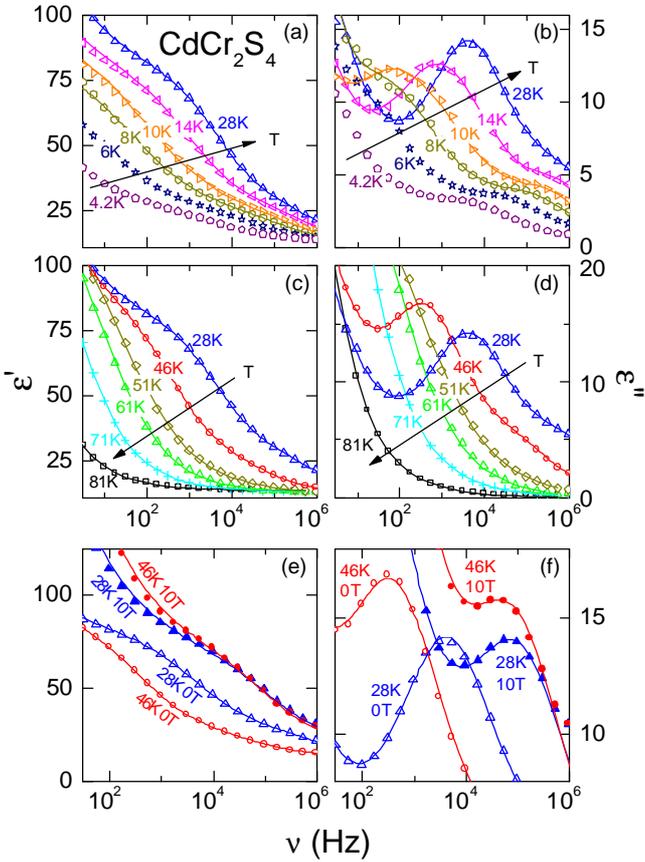

FIG. 3. Frequency dependence of $\varepsilon'$ (a,c,e) and $\varepsilon''$ (b,d,f) of CdCr$_2$S$_4$ for selected temperatures at $T < T_c$. Figs. 3(a-d) give the results without magnetic field and Fig. 3(e,f) shows examples for the influence of a field of 10 T on the relaxation dynamics. The lines are least-square fits as described in the text.

The most important outcome of this analysis is the quantitative information on $\tau(T)$, which was complemented by reading off loss-peak positions in the frequency and temperature-dependent plots [indicated, e.g., by the arrows in Fig. 1(b)]. In Fig. 4, the resulting temperature dependence of $\tau$ is shown in an Arrhenius representation for magnetic fields of 0, 5, and 10 T. Without magnetic field, at temperatures well above the magnetic phase transition ($100/T_c \approx 1.2$ K$^{-1}$), thermally activated behavior, $\tau = \tau_0 \exp(E/T)$, with an energy barrier $E$ of about 330 meV is observed (inset of Fig. 4). However, close to $T_c$ the temperature dependence of $\tau$ becomes reversed and it decreases by six orders of magnitude while the magnetic order develops. Finally, after reaching a minimum close to 28 K ($100/T \approx 3.6$ K$^{-1}$), $\tau(T)$ increases again when, with the further reduction of $T$, the decreasing thermal energy once more becomes the dominating factor. The slope in this region is rather small, leading to an unreasonably low attempt frequency, $1/(2\pi\tau_0) \approx 10^5$ Hz, and thus other mechanisms as, e.g., tunneling processes may play a role here.

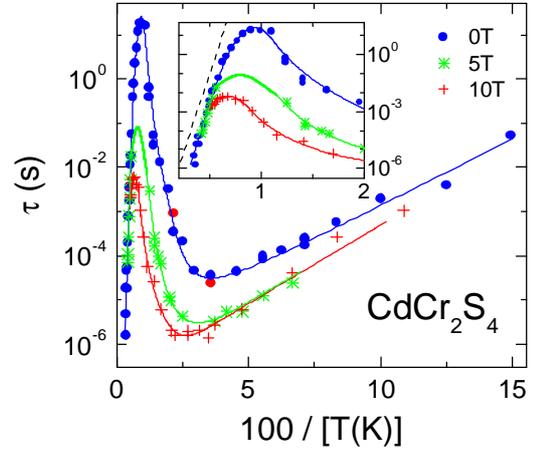

FIG. 4. Arrhenius plot of the relaxation time of CdCr$_2$S$_4$ without magnetic field and with fields of 5 and 10 T. The lines are drawn to guide the eyes. The inset gives a magnified view of the region around $T_c$. The dashed line indicates thermally activated behavior with an energy barrier of 330 meV.

Within this scenario the two main peaks observed in the temperature-dependent conductivity [arrows in Fig. 1(b)], can be explained as follows, considering for example the curve at 9.5 Hz in Fig. 1(b): Coming from high temperatures, the relaxation time $\tau$ of the relaxing entities grows until at about 170 K $1/(2\pi\tau)$ becomes equal to the frequency of the exciting field. Then a peak shows up in $\sigma'(T)$ as revealed by the closed squares in Fig. 1(b). When under further cooling below $T_c$ the ferromagnetic order rises, the relaxation speeds up again and for a second time the condition $1/(2\pi\tau) = \nu$ is fulfilled, which for 9.5 Hz is the case at about 65 K, leading to a peak in $\sigma'(T)$ accompanied by the strong increase of $\varepsilon'$ below $T_c$. Under further cooling, the order parameter saturates and again the decreasing thermal energy governs $\tau(T)$, finally leading to the peak (and decrease of $\varepsilon'$) observed at about 7 K.

As seen in Fig. 4, under application of a magnetic field the reversal of $\tau(T)$ already occurs at higher temperatures, as the system is driven towards ferromagnetism leading to a rise of the magnetization already at higher $T$. While the relative reduction of $t(T)$ below $T_c$ is less than for $H = 0$, for 10 T the relaxation time drops to a value of $10^{-6}$ s, more than a decade lower than without magnetic field. Both, the higher-temperature onset of the reduction and the shorter relaxation time finally reached, contribute to the observed colossal MC effect. While the upturns of $e'$ and $e''$ with decreasing $T$ in Fig. 2 can be explained in this way, it remains to be clarified why $e'$ reaches that much higher peak values for higher fields. Similar to canonical ferroelectrics, the static dielectric constant in $CdCr_2S_4$ is expected to exhibit a maximum close to the Curie-Weiss temperature of 130 K. As the upturns of $e'$ occur closer to this peak for higher magnetic fields, this explains the increasing peak values of $e'$.

While the results presented in this letter strongly suggest that it is the acceleration of the relaxation dynamics below $T_c$ and under application of an external magnetic field that leads to the colossal MC effects in $CdCr_2S_4$, it remains to be clarified what is the microscopic origin of the detected relaxation dynamics and why this dynamics couples so strongly to the magnetic order parameter. A coupling via exchangestriction, i.e. volume changes arising from the magnetic exchange energy, was proposed in ref. [2]. Exchangestriction in $CdCr_2S_4$ was already considered in earlier works in the discussion of the optical properties [7]. It should soften the lattice, thereby reducing the energy barriers against dipolar reorientation and thus enhancing the mean relaxation rate. As an alternative explanation of the MC effect one could consider a magnetic-field induced variation of charge-carrier mobility or density. A sizable dc magnetoresistance effect is well known for $CdCr_2S_4$ [8], but it cannot be responsible for the observed anomalies of $e'$, the dc resistivity only contributing to $e''$. In contrast, hopping-type charge transport is known to give rise to ac conductivity, approximated by the UDR, which via the Kramers-Kronig relation leads to $e' \sim n^{s-1}$ [6]. As a significant UDR contribution was indeed found in our analysis of the loss, an explanation of the detected anomalies of $e'$ in terms of a magnetoimpedance seems possible. However, this scenario is at odds with the loss peaks and steps in $e'$ observed in the magnetically ordered state, whose presence clearly proves that relaxational behavior prevails in this region. As becomes obvious in Fig. 3, only at the lowest frequencies the enhancement of hopping conductivity by magnetic order plays some role, constituting a background for the relaxational features in the spectra.

Finally, one should consider the possibility that the observed relaxation features are not a bulk property, but of the so-called Maxwell-Wagner type [9], i.e. caused by polarization effects at or close to the surface of the samples. A prominent example for Maxwell-Wagner relaxations are the so-called "colossal dielectric constant" materials [10,11]. In most (if not all) of these materials, relaxational behavior with very high (typically $10^3 - 10^5$), non-intrinsic dielectric constants arises from the formation of depletion layers at the interface between sample and metallic contacts [11]. The high capacitance of these insulating layers at the sample surface can lead to apparently high dielectric constants and relaxational behavior. However, in the present case, via measurements with different contact materials, such a scenario could be excluded [2]. But one should note that in [12] it was speculated that in sulfur-deficient $CdCr_2S_4$ a capacitive layer may form due to magnetic disorder at the surface and indeed the observation of a surface-related MC effect was reported. However, it is questionable if these results on highly doped $CdCr_2S_4$, with its several orders of magnitude higher conductivity, are of relevance for the present results on the undoped material.

In conclusion, by performing a thorough investigation of the relaxational behavior in $CdCr_2S_4$, we have provided strong evidence that the variation of the dielectric constant at the magnetic transition and the colossal MC effect in this material are caused by a speeding up of relaxation dynamics under the formation of magnetic order. Clearly the present dielectric experiments cannot provide final evidence on the microscopic origin of this puzzling behavior of $CdCr_2S_4$. While hopping charge transport or contact effects seem unlikely, the question if a bulk relaxation mechanism interacting with magnetic order via exchangestriction or an intrinsic surface-related Maxwell-Wagner relaxation causes this behavior, remains to be clarified.

This work was partly supported by the Deutsche Forschungsgemeinschaft via the Sonderforschungsbereich 484 and partly by the BMBF via VDI/EKM.

[1] T. Kimura *et al.*, Nature **426**, 55 (2003); N. Hur *et al.*, *ibid.* **429,** 392 (2004); Th. Lottermoser *et al.*, *ibid.* **430**, 541 (2004); T. Goto *et al.*, Phys. Rev. Lett. **92,** 257201 (2004); N. Hur *et al.*, *ibid.* **93**, 107207 (2004).
[2] J. Hemberger, P. Lunkenheimer, R. Fichtl, H.-A. Krug von Nidda, V. Tsurkan, and A. Loidl, Nature **434**, 364 (2005).
[3] P. Lunkenheimer *et al.*, Contemp. Phys. **41**, 15 (2000).
[4] The further increase of $e'$ towards higher temperatures is due to contact and conductivity contributions (see [2]).
[5] G. A. Samara, J. Phys. Cond. Matt. **15**, R367 (2003); L. E. Cross, Ferroelectrics **76**, 241 (1987).
[6] A. K. Jonscher, *Dielectric Relaxations in Solids* (Chelsea Dielectrics Press, London, 1983).
[7] G. W. Martin, A. T. Kellog, R. L. White, and R. M. White, J. Appl. Phys. **40**, 1015 (1969); E. Callen, Phys. Rev. Lett. **20**, 1045 (1968).
[8] H. W. Lehmann and M. Robbins, J. Appl. Phys. **37**, 1389 (1966).
[9] J. C. Maxwell, *Treatise on Electricity and Magnetism,* 3rd ed. (Dover, New York, 1991); R. J. Wagner, Ann. Phys. (Leipzig) **40**, 817 (1913).
[10] see, e.g., C. C. Homes *et al.*, Science **293**, 673 (2001); J. Wu, C. W. Nan, and Y. Deng, Phys. Rev. Lett. **89**, 217601 (2002); T. Park *et al.*, *ibid.* **94**, 017002 (2005).
[11] P. Lunkenheimer *et al.*, Phys. Rev. B **66**, 052105 (2002); **70**, 172102 (2005).
[12] M. Toda, *Proceeding of the 3rd Conf. on Solid State Devices, Tokyo, 1971*, Supplement to Oyo Buturi **41**, 183 (1971).